\documentclass[preprint,amsmath,amssymb,showkeys]{revtex4-1}
\usepackage{graphicx}
\usepackage{enumerate}
\usepackage{longtable}
\usepackage{xcolor}
\usepackage{url}

\begin{document}

\title{Tracing contacts to evaluate the transmission of COVID-19 from highly exposed individuals in public transportation}

\author{Caio Ponte$^{1}$,
Humberto A. Carmona$^{2}$,
Erneson A. Oliveira$^{1,3,4*}$\footnote[0]{$^*$Correspondence to: erneson@unifor.br},
Carlos Caminha$^{1}$,
Antonio S. Lima Neto$^{5,6}$,
José S. Andrade Jr.$^{2}$,
Vasco Furtado$^{1,7}$}

\affiliation{$^1$ Programa de Pós Graduação em Informática Aplicada, Universidade de Fortaleza, 60811-905 Fortaleza, Ceará, Brasil.\\
$^2$ Departamento de Física, Universidade Federal do Ceará, 60455-760 Fortaleza, Ceará, Brasil.\\
$^3$ Laboratório de Ciência de Dados e Inteligência Artificial, Universidade de Fortaleza, 60811-905 Fortaleza, Ceará, Brasil.\\
$^4$ Mestrado Profissional em Ciências da Cidade, Universidade de Fortaleza, 60811-905, Fortaleza, Ceará, Brazil.\\
$^5$ Célula de Vigilância Epidemiológica, Secretaria Municipal da Saúde, 60810-670 Fortaleza, Ceará, Brasil.\\
$^6$ Departamento de Saúde Pública, Universidade de Fortaleza, 60451-970 Fortaleza, Ceará, Brasil.\\
$^7$ Empresa de Tecnologia da Informação do Ceará, Governo do Estado do Ceará, 60130-240 Fortaleza, Ceará, Brasil.}

\date{\today}

\begin{abstract}
{\bf We investigate, through a data-driven contact tracing model, the transmission of COVID-19 inside buses during distinct phases of the pandemic in a large Brazilian city. From this microscopic approach, we recover the networks of close contacts within consecutive time windows. A longitudinal comparison is then performed by upscaling the traced contacts with the transmission computed from a mean-field compartmental model for the entire city. Our results show that the effective reproduction numbers inside the buses, $Re^{bus}$, and in the city, $Re^{city}$, followed a compatible behavior during the first wave of the local outbreak. Moreover, by distinguishing the close contacts of healthcare workers in the buses, we discovered that their transmission, $Re^{health}$, during the same period, was systematically higher than $Re^{bus}$. This result reinforces the need for special public transportation policies for highly exposed groups of people.}
\end{abstract}
\maketitle

\section{Introduction}

Human mobility is crucial to understanding the COVID-19 pandemic since the Severe Acute Respiratory Syndrome Coronavirus 2 (SARS-CoV-2) is disseminated individual to individual via droplet and airborne transmissions~\cite{thelancet2020}. Considering that non-pharmacological interventions, such as social distancing and isolation, still represent fundamental measures to control the COVID-19 outbreak, the nature of SARS-CoV-2 dissemination also unveils the need to understand the role of the space where interaction between people occurs. There is consensus that superspreading events, which are usually investigated through contact tracing models \cite{lloyd-smith2005, lossio-ventura2020, sefarino2021, reyna-lara2021, hamner2020, majra2021}, are more likely to happen in indoor environment, as substantiated by previous studies of indoor contagion in hospitals~\cite{liu2020, lednicky2020}, restaurants~\cite{kwon2020}, offices~\cite{bohmer2020}, and even on cruise ships~\cite{kakimoto2020, mizumoto2020}. However, the relation between the microscopic level of contagion in indoor environments and the macroscopic observables, such as the numbers of cases and deaths at the city scale remains unclear.

Public transportation is one of the main forms of commuting, playing an important role in the pace of life in cities~\cite{vuchic2017}, specially in epidemics \cite{stoddard2009, edelson2011, bomfim2019, kraemer2020, schlosser2020, melo2021}. In spite of the fact that some cities have adopted social distancing and sanitary protocols on public transportation to control the COVID-19 outbreak, it is common that buses or subways get crowded at rush hour, mainly in the developing countries. In recent months, some studies have been proposed to establish the safety of public transportation regarding the indoor COVID-19 contagion~\cite{dicarlo2020, gosce2020, jenelius2020, shen2020a, shen2020b, zhang2021, hu2021}. To the best of our knowledge, however, none of these studies have considered the possibility of a comparative analysis based on a two-fold perspective, namely, the dynamic of people's movement in a city and the dynamic of the virus dissemination within vehicles of public transport.

Here, using data about people's movement on buses and COVID-19 infection in a large metropolis, we define two data-driven mathematical models based on concepts of complex networks and non-linear dynamics in order to foster the understanding of the role public transportation plays in the COVID-19 pandemic. At the microscopic scale, we define a contact tracing model to estimate the transmission within city buses and, at the macroscopic scale, a compartmental model is employed to estimate the transmission in the entire city. The main contribution of our study is a comparative analysis between these two distinct modeling approaches through the combination of daily epidemiological and mobility data during the first 9 months of the local COVID-19 outbreak, and through different social distancing restriction regimes. One specially relevant aspect of this work is the fact that we are able to trace within the public transportation vehicles (\emph{i.e.}, indoor environments) two groups of people, one of them with a higher exposure to the virus in comparison to the other. This allows us to shed light on potentially dangerous superspreading events in public transportation.

\section{Modeling Approach}

\subsection{Contact tracing model}

\begin{figure}[!htb]
\centering
\includegraphics[width=0.5\textwidth]{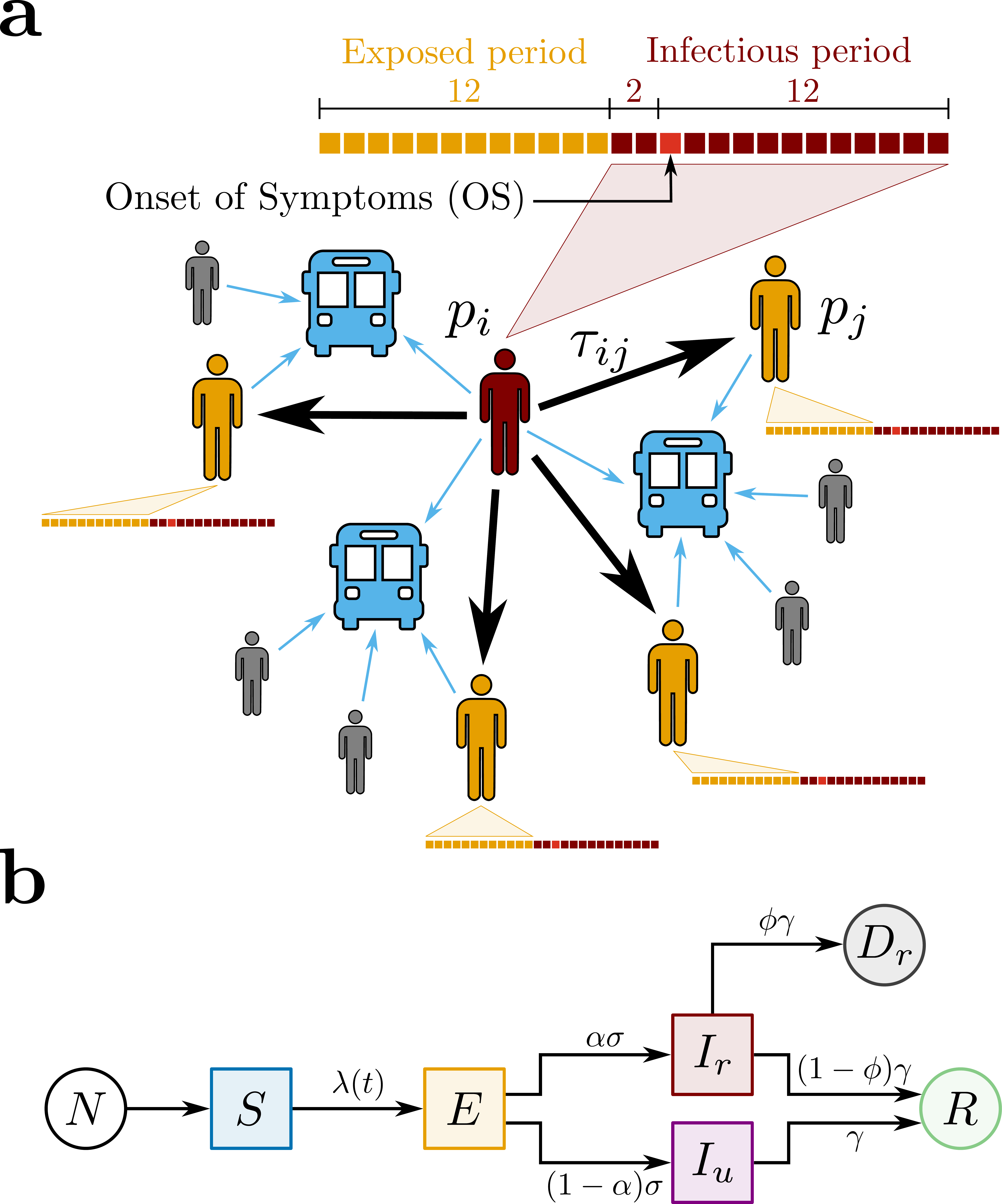}
\caption{{\bf Proposed Models for COVID-19 and Spreading Scenarios.} {\bf (a)} Potentially Infectious Contacts (PICs). We define a PIC when an infectious passenger $p_i$ (in red) and an exposed passenger $p_j$ (in yellow) share the same bus. The weight $\tau_{ij}$ is the estimated value of the ride time shared by $p_i$ and $p_j$. The time lines show the infectious (in red) and the exposed periods (in yellow) of each passenger, where each square represents one day. The time lines are built based on the Onset of Symptoms (OS). Precisely, the infectious period begins 2 days before OS and ends 12 days after OS, while the exposed period begins 14 days before OS and ends 2 days before OS. Other passengers (in gray), even though they have shared the same bus with $p_i$, either were not notified as COVID-19 cases or, however notified, they were not considered as PICs because they were not in their exposed period. {\bf (b)} The SEIIR model. The total population of size $N$ provides the susceptible population $S$ (in blue). The susceptible individuals become exposed $E$ (in yellow) at a time-dependent rate $\lambda(t)$. The exposed individuals become infectious at a time rate $\sigma$. A fraction $\alpha$ of the infectious population is reported $I_r$ (in red), while a fraction $(1-\alpha)$ is unreported $I_u$ (in purple). The infectious individuals that recover, reported or not, become recovered $R$ (in green) at a time rate $\gamma$. Finally, it is assumed that a fraction $\phi$ of the removed population $\gamma I_{r}$ deceases $D_{r}$ (in dark gray).}
\label{fig:models}
\end{figure}

We propose a contact tracing model using two datasets that relate bus validations to COVID-19 confirmed cases during the periods of social isolation, lockdown, and economic reopening in the city of Fortaleza, Cear\'a, Brazil (see Methods). Our model is a network based on Potentially Infectious Contacts (PICs), in which bus passengers during their infectious period - according to subsequent diagnosis of COVID-19 - have shared the transport for a certain amount of time with other passengers, the latter in their exposed period - also according to subsequent COVID-19 diagnosis. Precisely, the proposed network is composed of vertices $p_i$ that represent the passengers diagnosed with COVID-19, and weighted directed edges $c_k=(p_i,p_j,\tau_{ij})$ that represent PICs. For each edge, the direction is assigned from an infectious passenger $p_i$ to an exposed passenger $p_j$, and the weight $\tau_{ij}$ is defined as the estimated value of the ride time shared by $p_i$ and $p_j$ on the same bus, as shown in Fig.~\ref{fig:models}a. We calculate $\tau_{ij}$ by superimposing the estimated ride times from $p_i$ and $p_j$, considering the different moments of their boarding. Here, the epidemiological profile for COVID-19 transmission is characterized by the dates of the passengers' Onset of Symptoms (OS). The infectious period corresponds to the days in which a passenger diagnosed with COVID-19 can transmit the virus, initiating 2 days before OS and ending 12 days after OS. The exposed period refers to the time window during which the passenger can get the virus and maintain it latent until the infectious period. In this context, the exposed period begins 14 days before OS and ends 2 days before OS,~\emph{i.e.}, the infectious and the exposed periods have a width of 14 and 12 days, respectively, and they do not overlap~\cite{he2020, li2020a, sanche2020}. Furthermore, if there is more than one PIC related to an exposed passenger $p_j$, we consider solely the edge with the largest value of $\tau_{ij}$. It is important to notice that, by crossing the datasets of bus validations and confirmed cases of COVID-19 in Fortaleza during the period from March to December 2020, we are able to identify $5,159$ pairs of infectious and exposed passengers that rode the same bus on the same day. However, their associated values of $\tau_{ij}$ could only be computed for $3,023$ ($58.6\%$), due to missing information in the dataset of bus validations. From these pairs, we obtain that the network of PICs corresponds to a forest composed of $213$ trees with a total of $530$ vertices (infectious passengers) and $317$ edges (PICs). From all vertices found, $97$ were identified as healthcare workers (see Methods). The Centers for Disease Control and Prevention (CDC) recommends that any contact tracing strategy for COVID-19 should consider the concept of Close Contacts (CCs)~\cite{cdc2020},~\emph{i.e.}, anybody who has been for at least $15$ minutes within $6$ feet ($\approx 2$ meters) of an infectious person. Since buses are small, enclosed, and they have a great tendency to get crowded at rush hours, we define the CCs in the network of PICs only considering the time condition $\tau_{ij}>\tau_c$, where the threshold $\tau_c=15$ minutes. Applying this criterion to the network of PICs, we find that the network of CCs is composed of $154$ trees with a total of $360$ vertices (infectious passengers) and $206$ edges (CCs). In this case, $75$ vertices were identified as healthcare workers. In order to understand the COVID-19 spreading in public transportation, we define the effective reproduction number for the contact tracing model, $Re^{bus}$, as the expected number of secondary cases produced by a single (typical) infection. Precisely, it accounts for two contributions in relation to who is spreading the disease: one due to reported infectious individuals, $Re_r^{bus}$, and another due to unreported infectious individuals $Re_u^{bus}$. Here, we assume that the fraction of newly reported to newly unreported cases generated by a typical reported infectious individual remains invariant during time. This is equivalent to consider the value of $Re_r^{bus}$ proportional to the average number of outdegrees from the vertices in the network of CCs during a given time window, $\langle d_{out}^{CCs} \rangle$,
\begin{equation}
Re_r^{bus}=\chi \langle d_{out}^{CCs} \rangle.\label{eq:outdegree}
\end{equation}
The constant of proportionality $\chi$ involved in this relation will be explicitly computed through the calibration between the contact tracing and the compartmental models. Each consecutive time window has a width of 22 days and a step size of 5 days. We emphasize that our model has an intrinsic time delay regarding the consolidation of $Re_r^{bus}$ that can reach $\approx 53$ days. This value is associated to the time delay in the consolidation of COVID-19 dataset ($\approx 15$ days) and to the superposition of the maxima of two infectious periods and one exposed period.

\subsection{Compartmental Model}

We also adopt a compartmental model to describe the transmission of COVID-19 in order to estimate the levels of infection of the pathogen in Fortaleza. Here, we propose a SEIIR model that distinguishes the populations of Susceptible, Exposed, Infectious (reported or unreported), and Removed (recovered or deceased) individuals, as shown in Fig.~\ref{fig:models}b. Our model is inspired by the SEIIR model proposed by Li~\emph{et al.}~\cite{li2020b}. The reported infectious population $I_r$ corresponds to the number of individuals that had the SARS-CoV-2 infection confirmed by the health system. The unreported infectious population $I_u$ comprises the complement of $I_r$,~\emph{i.e.}, individuals that were infected with COVID-19 but remained unknown to health authorities. We assume that the large majority of the reported infectious individuals are symptomatic cases, in contrast to the population of unreported infectious individuals - of which the large majority is assumed to be of asymptomatic cases. Given this fundamental assumption and considering the recent finding that asymptomatic people are 42\% less likely to transmit the SARS-CoV-2 than symptomatic ones~\cite{byambasuren2020}, we define that the transmission rate for the unreported infectious population $I_u$ is reduced by a factor of $\mu$ in relation to the parameter $\beta$ that represents the transmission rate for the reported infectious population $I_r$. In this context, the time-dependent rate at which the susceptible population $S$ becomes the exposed population $E$ is given by
\begin{equation}
\lambda(t) = \beta \frac{\left(I_r + \mu I_u\right)}{N},\label{eq:force}
\end{equation}
where $N$ is the total population of Fortaleza, taken as constant, being approximately equal to $2.67$ million people. A fraction $\alpha$ of the exposed individuals is presumed to become reported infectious at a rate $\sigma$, and the complementary fraction $(1-\alpha)$ to evolve to unreported infected at the same rate. Also, both reported and unreported infectious population are assumed to become part of the removed population at the same rate $\gamma$. We also keep track of the fraction $\phi$ of the removed reported infectious population evolving to death, so that the reported deceased population $D_r$ increases at a rate of $\phi \gamma I_r$. The following system of coupled differential equations rules our model:
\begin{align}
    \frac{dS}{dt} &= -\lambda S,\label{subeq:model1}\\
    \frac{dE}{dt} &= \lambda S - \sigma E,\label{subeq:model2}\\
    \frac{dI_r}{dt} &= \alpha \sigma E - \gamma I_r,\label{subeq:model3}\\
    \frac{dI_u}{dt} &= \left( 1- \alpha\right) \sigma E - \gamma I_u,\label{subeq:model4}\\
    \frac{dR}{dt} &= \left( 1- \phi\right) \gamma I_r + \gamma I_u,\label{subeq:model5}\\
    \frac{dD_r}{dt} &= \phi \gamma I_r.\label{subeq:model6}
\end{align}
The total population $N = S+E+I_r+I_u+R+D_r$ is conserved. Furthermore, it can be readily shown~\cite{li2020b} that the effective reproduction number $Re^{city}$ is given by
\begin{equation}
    Re^{city} = \left[\alpha\frac{\beta}{\gamma}+(1-\alpha)\mu\frac{\beta}{\gamma}\right]\frac{S}{N}.
    \label{eq:Re}
\end{equation}
From Eq.~\eqref{eq:Re}, we can identify $Re_r^{city}=(\beta/\gamma)(S/N)$ as the average number of secondary infections due to contagion with reported infectious individuals, while $Re_u^{city}=\mu (\beta/\gamma)(S/N)$ is the effective reproduction number due to contagion with unreported infectious individuals. Finally, the SEIIR model is used here as a core model within the Iterative Ensemble Kalman Filter (IEnKF) framework (see Methods). This approach allows us to investigate the time evolution of the effective reproduction number $Re^{city}$ by inferring the SEIIR model parameters and their populations (see Figs.~S1 and S2 of the Supplementary Information). The IEnKF framework is systematically applied to running windows of 22 days, with step size of 5 days, starting from March 24 to November 9, 2020. We use as observable the cumulative number of deaths by SARS-CoV-2 reported daily by the health authorities. For the first and subsequent windows, the guesses for the initial populations of exposed, $E^0$, and deceased individuals, $D^0$, are obtained from the daily number of COVID-19 confirmed cases and the daily cumulative number of COVID-19 confirmed deaths. As the cumulative number of deaths by SARS-CoV-2, both quantities are calculated from the dataset of COVID-19 confirmed cases and deaths (see Methods). In the particular case of the initial guess for the exposed population, $E^0=C_i/(\alpha\sigma)$, where $C_i$ is the reported number of daily cases. This corresponds, for example, to $4,982$ individuals in the first window. After using IEnKF to estimate the values of all model parameters for the first window, the factor $Re^{city}$ is calculated at its center. These parameters and all populations obtained by numerical integration of Eqs.~\eqref{subeq:model1}-\eqref{subeq:model6}, except for $E^0$ and $D^0$, as previously explained, are used as initial guesses for the second window. The same procedure is then repeated for the third and subsequent windows.

\section{Results and Discussion}

\begin{figure}[!htb]
\centering
\includegraphics[width=1.0\textwidth]{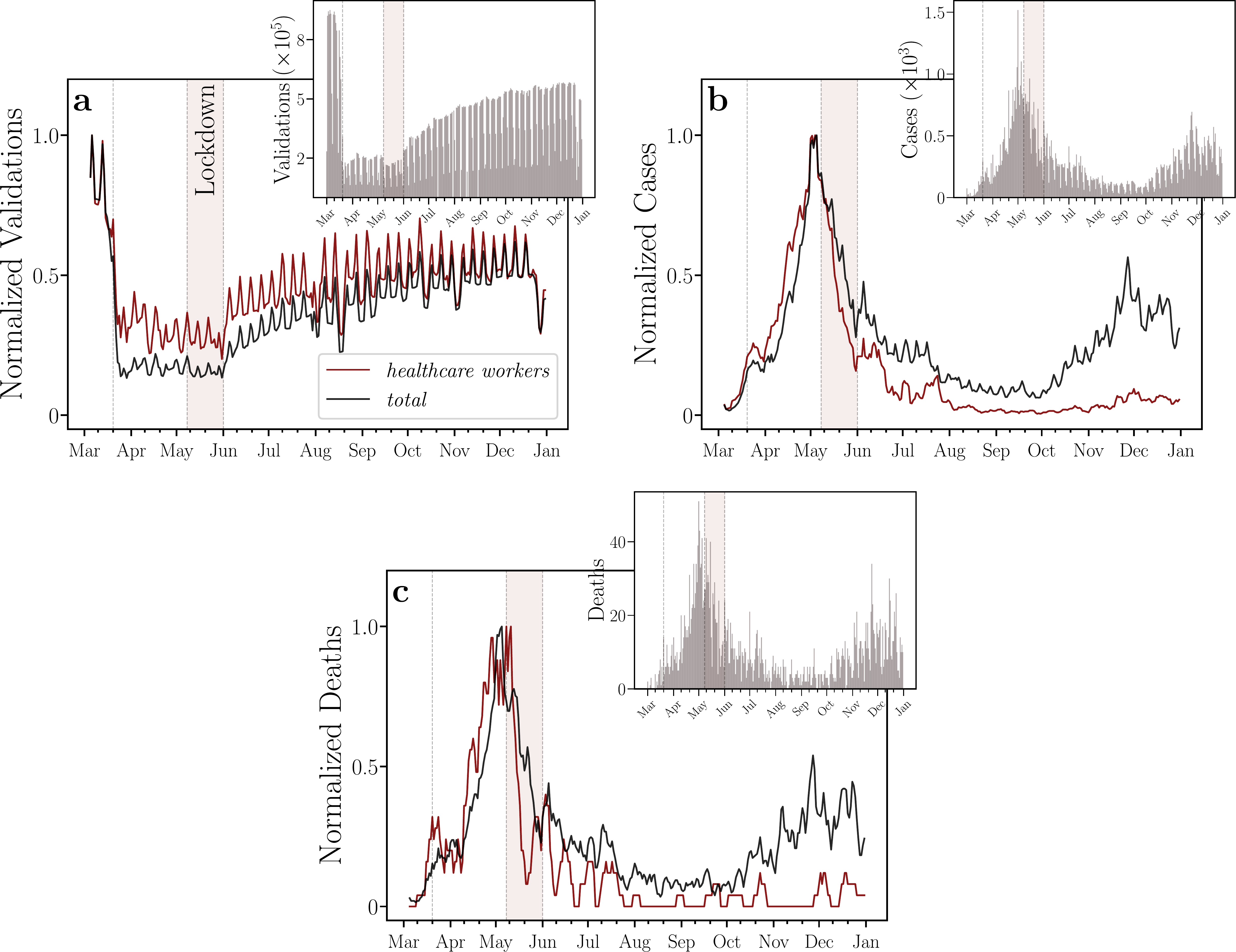}
\caption{{\bf Time series and moving averages of bus validations, COVID-19 cases, and COVID-19 deaths.} Time evolutions of the normalized moving averages of {\bf(a)} bus validations, {\bf(b)} COVID-19 cases, and {\bf(c)} deaths, for healthcare workers (in red) and all individuals (in black). The insets show their corresponding daily numbers. In {\bf(a)}, we note that healthcare workers that came into contact with SARS-CoV-2 during the studied period did not reduce their bus rides as much as other passengers. In addition, the normalized moving averages of bus validations of healthcare workers and of all individuals are getting closer to each other again as the economic reopening progresses. In {\bf(b)} and {\bf(c)}, we find that both the normalized moving averages of cases and of deaths, respectively, for healthcare workers increased before those of all individuals until the lockdown regime. The windows of moving averages have 5 days of width for all curves. We normalized each moving average by its maximum. The vertical dotted lines represent the beginning of social isolation (State Decree 33,519), lockdown (State Decree 33,574), and economic reopening (State Decree 33,608) regimes imposed on March 20, May 8, and June 1, 2020, respectively. We also highlight, in light red, the lockdown period in the city of Fortaleza.}
\label{fig:datasets}
\end{figure}

Figure~\ref{fig:datasets}a shows the normalized moving averages of bus validations of individuals that got COVID-19, including healthcare workers. We note that healthcare workers that came into contact with SARS-CoV-2 during the studied period did not reduce their bus rides as much as other passengers. In addition, their normalized moving averages of bus validations are getting closer to each other again as the economic reopening progresses. The inset of Fig.~\ref{fig:datasets}a shows the daily bus validations, which gradually started to increase in the economic reopening. Figures~\ref{fig:datasets}b and \ref{fig:datasets}c show the daily numbers of cases and deaths, respectively, following the same previous normalization and stratification.

\begin{figure}[!htb]
\centering
\includegraphics[width=0.85\textwidth]{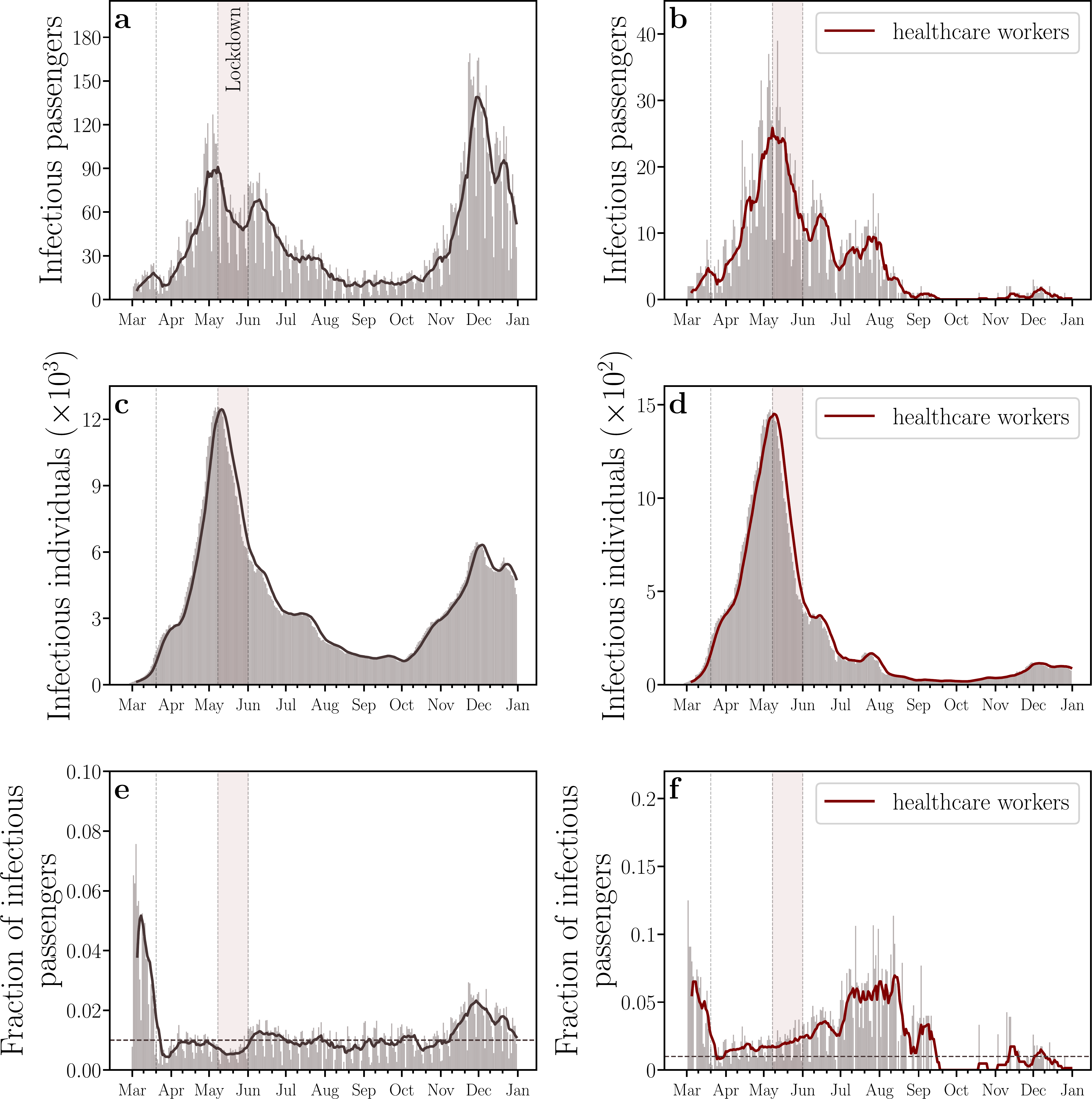}
\caption{{\bf Representativeness of the dataset of COVID-19 confirmed cases on buses.} The daily numbers of {\bf (a)} all infectious passengers, {\bf (b)} infectious passengers who are healthcare workers, {\bf (c)} all infectious individuals in the entire city, and {\bf(d)} infectious individuals who are healthcare workers in the entire city. For healthcare workers, we highlight an unexpected emergence of a single peak in the daily numbers of infectious individuals within buses and in the entire city, which contrasts to the first and the second waves of COVID-19. We conjecture that the explanation for this behavior may be the lack of Personal Protective Equipment (PPE) in hospitals during the first wave or the herd immunity of healthcare workers during the second wave. {\bf (e)} The fraction of all infectious passengers. {\bf (f)} The fraction of infectious passengers who are healthcare workers. These results show that the percentage of infectious passengers with respect to all infectious individuals in Fortaleza was higher than $1\%$ during most of the epidemic period (dashed gray line). All solid lines represent moving averages with windows of 7 days.}
\label{fig:representativeness}
\end{figure}

The representativeness of the dataset of COVID-19 confirmed cases on buses is assessed comparing the daily numbers of infectious individuals within those vehicles and in the entire city, as shown in Fig.~\ref{fig:representativeness}. The daily number of infectious individuals is computed taking into account the 14 days that the individuals remain infectious,~\emph{i.e.}, each individual who tested positive for COVID-19 counts up to 14 times, once per day, for the infectious curve. Figures~\ref{fig:representativeness}a and \ref{fig:representativeness}b show the daily numbers regarding all infectious passengers and those infectious passengers who are healthcare workers, respectively. Similarly, the daily numbers of infectious individuals and infectious healthcare workers of the entire city are shown in Figs.~\ref{fig:representativeness}c and \ref{fig:representativeness}d, respectively. While the first and second waves of the epidemic can be clearly identified in both curves shown in Fig.~\ref{fig:representativeness}a (infectious passengers) and Fig.\ref{fig:representativeness}c (infectious individuals), only highly attenuated peaks during the second wave period can be visualized in the corresponding curves for healthcare workers, as shown in Figs.~\ref{fig:representativeness}b and \ref{fig:representativeness}d. We conjecture that the explanation for this behavior may be twofold. First, due to the high contagion of healthcare workers during the first wave, this group of people may have achieved a large percentage of immunity, as compared to the rest of the population. Second, efficient Personal Protective Equipment (PPE) became more available in hospitals after the first wave. The results in Fig.~\ref{fig:representativeness}e show that the percentage of infectious passengers with respect to all infectious individuals in Fortaleza was higher than $1\%$ during most of the epidemic period. Finally, the evolution in time of the fraction between infectious passengers and infectious individuals in the city who are both healthcare workers is shown in Fig.~\ref{fig:representativeness}f.

\begin{figure}[!htb]
\centering
\includegraphics[width=1.0\textwidth]{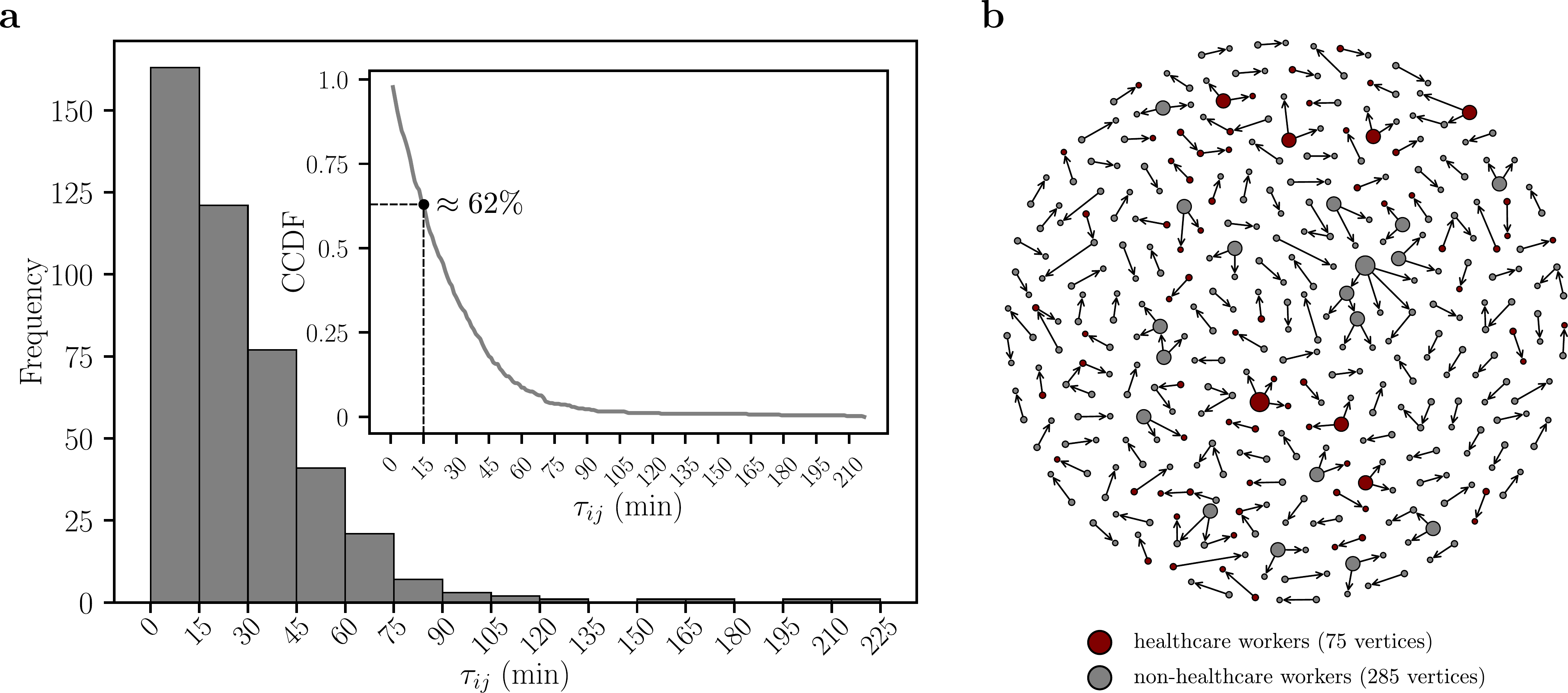}
\caption{{\bf Shared Ride Time Histogram $\tau_{ij}$ and Network of Close Contacts (CCs).} ({\bf a}) We show the weight distribution $\tau_{ij}$ of the network of Potentially Infectious Contacts (PICs) in 15-minute-width bins. The average of the shared ride times of PICs is $\langle\tau_{ij}\rangle^{PICs} \approx 28$ minutes. Applying the threshold $\tau_c=15$ minutes in the network of PICs, we define a network of Close Contacts (CCs). For the network of CCs, the average of the shared ride times is $\langle\tau_{ij}\rangle^{CCs} \approx 39$ minutes. The inset shows the Complementary Cumulative Distribution Function (CCDF) of $\tau_{ij}$. We find that the percentage of the edges with $\tau_{ij}$ greater than $\tau_c=15$ minutes is $\approx 62 \%$ (dashed line), {\it i.e.}, most part of PICs are CCs. ({\bf b}) The vertices represent the bus passengers that were diagnosed with COVID-19 and the edges corresponds to the CCs each passenger had using the public transportation system in Fortaleza. The healthcare and non-healthcare workers are represented by red and gray vertices, respectively. The size of the vertices is proportional to their outdegrees.}
\label{fig:histogram_network}
\end{figure}

The histogram of the values of $\tau_{ij}$ for the network of PICs is shown in Fig.~\ref{fig:histogram_network}a. The obtained distribution is characterized by the average $\langle \tau_{ij} \rangle^{PICs} \approx 28$ minutes. We find that CCs, defined by $\tau_{ij}>\tau_{c}=15$ minutes, represent about 62\% of the PICs, as shown by the Complementary Cumulative Distribution Function (CCDF) in the inset of Fig.~\ref{fig:histogram_network}a. For the network of CCs, the average of the shared ride times is $\langle\tau_{ij}\rangle^{CCs} \approx 39$ minutes. Fig.~\ref{fig:histogram_network}b shows the network of CCs taking into consideration the periods of social isolation, lockdown, and economic reopening. As depicted, it is composed of several trees, where the vertices represent bus passengers that were diagnosed with COVID-19 and the edges correspond to CCs. Bus passengers identified as healthcare workers in the network are highlighted in red. The size of the vertices is proportional to their outdegrees.

At this point, we show that it is possible to perform a direct comparison between the computed values of $Re_{r}^{bus}$ obtained from the contact tracing model for different time windows and the corresponding effective reproduction numbers $Re^{city}$ estimated from the compartmental model. First, it is reasonable to assume that $Re_{r}^{bus}=Re_{r}^{city }$, as long as the population traveling by public buses can be considered as statistically equivalent, from an epidemiologic point of view, to the rest of the city. As a consequence of this assumption and using Eq.~\eqref{eq:Re}, we can write that
\begin{equation}
    Re_r^{bus} = \psi Re^{city},
    \label{eq:Rfit}
\end{equation}
where the parameter $\psi=[\alpha+(1-\alpha)\mu]^{-1}$ depends on the time window used for model inference with the IEnKF technique. We now proceed with the comparison between contact tracing and compartmental models. In practical terms, this is achieved by upscaling $\langle d_{out}^{CCs} \rangle$ to the numerical values obtained for $Re^{city}$ during the early period of the SARS-CoV-2 epidemic, before the restrictions of isolation and social distancing imposed by the State Government took effect. Considering Eqs.~\eqref{eq:outdegree} and \eqref{eq:Rfit}, we use the relation $\chi \langle d_{out}^{CCs} \rangle = \psi Re^{city}$ and the numerical values of $\langle d_{out}^{CCs} \rangle$, $\psi$ and $Re^{city}$ on the day that corresponds to the maximum of $Re^{city}$ during the first wave (April 8, 2020) to calculate $\chi\approx 37$. This constant combined with the values of $\psi$ from the inference with the compartmental model, and the values of $\langle d_{out}^{CCs} \rangle$ from the contact tracing, both calculated for all time windows, are then used to obtain the entire curve of $Re^{bus}=(\chi/\psi) \langle d_{out}^{CCs} \rangle$ (see the time evolution of $\chi/\psi$ in Fig.~S3 of the Supplementary Information). The value of $\chi$ can be understood as the product of two factors, $\chi=\chi_{rr}\chi_{ru}$. Assuming the equality between the proportions of CCs in the pairs of infectious and exposed passengers with existing and missing values of $\tau_{ij}$, we estimate $\chi_{rr} \approx 1/0.586 \approx 1.70$ as a balance factor for possible missing CCs. The value of the remaining factor $\chi_{ru} \approx 21.76$ expresses the sub-notification of the confirmed cases as well as our lack of knowledge on the transmission from reported to unreported infectious passengers, for which the factor $(1-\alpha)/\alpha$ could be a lower bound (see Fig.~S2 of the Supplementary Information for a sensitivity analysis of the model with parameter $\alpha$), approximately between 5 and 9 \cite{li2020b}. In an entirely similar fashion, by considering only reported infectious passengers that can be identified as healthcare workers, we can estimate their particular effective reproduction number as $Re^{health}=(\chi/\psi)\langle d_{out}^{CCs} \rangle^{health}$ with the same upscaling factor $\chi/\psi$ used for all infectious passengers and $\langle d_{out}^{CCs} \rangle^{health}$ is the average of the vertices outdegrees for healthcare workers.

\begin{figure}[!htb]
\centering
\includegraphics[width=0.9\textwidth]{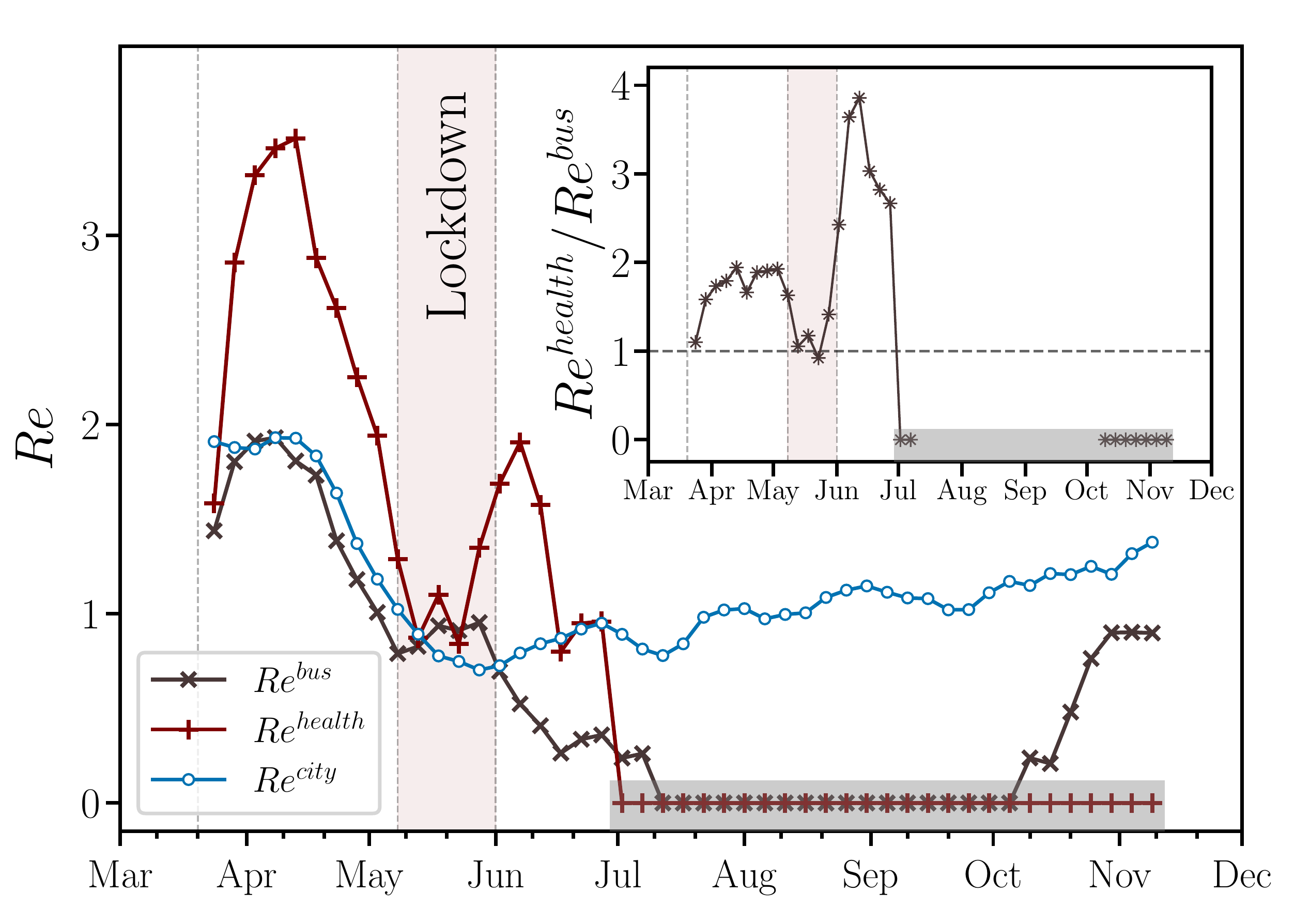}
\caption{{\bf Time evolution of the effective reproduction numbers.} Moving averages of the effective reproduction number for the entire city, $Re^{city}$ (blue $\circ$), for buses, $Re^{bus}$ (gray $\times$), and for healthcare workers in the buses, $Re^{health}$ (red $+$). We find that $Re^{bus}$ consistently follows $Re^{city}$ during the local COVID-19 outbreak, except for a three-month period between the first and the second waves of daily cases. We also show that $Re^{health}$ was systematically higher than $Re^{bus}$, which unveils that the healthcare workers played an important role in the transmission within buses during the first wave of COVID-19 in Fortaleza. The inset shows that the maximum ratio $Re^{health}/Re^{bus}$ occurred soon after the lockdown period. The windows of moving averages have 22 days of width with step size of 5 days for all curves. In the period indicated by the shaded regions in the main plot and its inset, both $Re^{bus}$ and $Re^{health}$ decayed to undetectable standards, {\it i.e.}, no CCs could be identified under the framework of our contact tracing approach. The vertical dotted lines represent the beginning of social isolation (State Decree 33,519), lockdown (State Decree 33,574), and economic reopening (State Decree 33,608) regimes imposed on March 20, May 8, and June 1, 2020, respectively. We also highlight, in light red, the lockdown period in the city of Fortaleza.}
\label{fig:Re}
\end{figure}

In Figure~\ref{fig:Re}, we show the comparison between the estimates of $Re^{bus}$ and $Re^{city}$ from March to November 2020. Although the contact tracing and compartmental models are defined on different scales, the former on a microscopic scale and the latter on a macroscopic scale, the two curves capture the same decreasing trend associated to both social isolation and lockdown periods. We note that $Re^{bus}$ consistently follows $Re^{city}$ during the local COVID-19 outbreak, except for a three-month period between the first and the second waves of daily cases. In this period, the $Re^{bus}$ decayed to undetectable standards despite the fact that the number of daily bus validations has increased (see Fig.~\ref{fig:datasets}a). As also shown in Fig.~\ref{fig:Re}, $Re^{health}$ was systematically higher than $Re^{bus}$, which unveils that the healthcare workers played an important role in the transmission within buses during the first wave of COVID-19 in Fortaleza. Furthermore, $Re^{health}$ remained undetectable even in the beginning of the second wave, in contrast to $Re^{bus}$ and notwithstanding the increase of the number of daily bus validations of healthcare workers, as shown in Fig.~\ref{fig:datasets}a. As shown in the inset of Fig.~\ref{fig:Re}, the maximum ratio $Re^{health}/Re^{bus}$ occurred soon after the lockdown period, since the hospitals were still overloaded due to the peak of cases at the beginning of May and the new daily infections were low in the beginning of the reopening period. We emphasize that the complement of $Re^{health}$, due to non-healthcare workers, behaves similar to $Re^{bus}$.

\section{Conclusions}

In summary, two epidemiological models have been used in this work to understand the transmission on public transportation during the COVID-19 outbreak in Fortaleza, Cear\'a, Brazil. Whilst the compartmental model accounts for the transmission in the entire city (macroscopic scale), the contact tracing model has been used to estimate the transmission inside city buses (microscopic scale) through the concept of CCs. Both models were fed with real data of bus validations and of COVID-19 confirmed cases and deaths. Our results show that $Re^{bus}$ consistently follows $Re^{city}$ during the local COVID-19 outbreak, except for a three-month period between the first and the second waves of daily cases. Furthermore, the transmission from healthcare workers within buses until the end of July is characterized by a value of $Re^{health}$ persistently greater than $Re^{bus}$. Healthcare workers, even the non-frontline professionals, are more likely to get and, consequently, spread the pathogen because their social network distances to individuals that tested positive for COVID-19 are very short compared to non-highly exposed workers. Despite being more tested, healthcare workers may not even know that they are infectious when they board a bus due to eventual time delays of the result of a COVID-19 test. Other groups of highly exposed people may affect the dynamics of dissemination of the virus in a similar way,~\emph{e.g.}, education workers and police officers. Therefore, our results reinforce the worldwide claim that it is imperative to propose special policies to support displacement (or to avoid it) of highly exposed groups of people. Finally, we suggest that the intensity and the necessity of using public transportation by highly exposed groups must be seriously considered as a criterion to prioritize their vaccination.

\section{Methods}

\subsection{Datasets}

\subsubsection{Bus validations}

Most part of bus passengers in Fortaleza ($\approx$ 94\%) pay their bus fares with a smart card. Every time a passenger passes their card on a ticket gate of a bus, a validation record is created. The Fortaleza City Hall compiled and made available an anonymized dataset of bus validations with the following information: a citizen's ID (a hash code), a vehicle ID (another hash code), the date and time of the validation record and the estimated ride time. The dataset ranges from March to December $2020$, totaling $107,488,528$ validation registers that refers to $1,426,569$ different passengers.

\subsubsection{COVID-19 confirmed cases and deaths}

The dataset of COVID-19 confirmed cases and deaths is an anonymized list of all individuals diagnosed with the disease in Fortaleza from March to December $2020$. These data were also processed and made available by the Fortaleza City Hall. Such dataset is organized in columns as follows: a citizen's ID (the same hash code used in the previous dataset), the date of OS, a confirmed death flag, the date of death and a healthcare worker flag. In the period of time ranged by the data, there are $85,553$ confirmed cases ($5,960$ of healthcare workers) and $3,075$ confirmed deaths ($227$ of healthcare workers). We emphasize that these healthcare workers are not only the frontline professionals but also people whose jobs are related to the health field. Finally, we found that $9,032$ people ($721$ healthcare workers) were diagnosed with COVID-19 and used their smart card on buses at least once from March to December $2020$.

\subsection{Iterated Ensemble Kalman Filter}

We use the Iterated Ensemble Kalman Filter (IEnKF) framework~\cite{ionides2006, li2020b, king2008, sakov2012} to infer the compartmental model parameters and initial subpopulations. The algorithm is based on comparing predictions of the model $f(.)$ obtained by the numerical integration of Eqs.~\eqref{subeq:model1}-\eqref{subeq:model5} of the main text with a set of $T$ observations $\mathcal{O}_1, \dots, \mathcal{O}_T$ taken at discrete times $t_1, \dots, t_T$ within an observation window (see Fig.~S4 of the Supplementary Information). The inference framework starts from an initial state vector $X^{(0)} = \left\{S,E,I_r,I_u, R, D_r \right\}^{(0)}$, and an initial parameter vector $\theta^{(0)} = \left\{\beta, \mu, \sigma, \gamma, \alpha, \phi\right\}^{(0)}$. To these vectors, uncertainties are attributed in terms of the variance matrices $\sigma_{X}$ and $\sigma_{\theta}$, respectively. For each iteration $m$, an ensemble of $P$ ``particles'' is generated such that each particle has the initial state at time $t_0$ drawn from a multivariate normal distribution with mean $X^{(m-1)}$ and variance $a^{(m-1)}\sigma_{X}$, where $0<a<1$ is a ``cooling factor''. The initial state vector for particle $i$ is denoted by $X(t_0,i) = \mathcal{N}( X^{(m-1)}, a^{(m-1)}\sigma_{X})$. These states are also used to set $X_F(t_0,i)$ which define the posterior distribution at time $t_0$. Analogously, each particle $i$ has an initial parameter vector $\theta(t_0,i) = \mathcal{N}(\theta^{(m-1)}, b^{(m-1)}\sigma_{\theta})$, where $0<b<1$ is another cooling factor. The inference proceeds by numerically integrating the model from these initial conditions, such that the predicted vector state for each particle $i$ at time $t_n$ is obtained from the following distribution, $X_P(t_n, i) = f(X_F(t_{n-1}, i), \theta(t_{n-1}, i))$. Based on these predictions, a weight $W(t_n, i)$ is assigned to each particle $i$, such that
\begin{equation}
    W(t_n, i) = \exp\left(-\frac{|\mathcal{O}(t_n,i) - \mathcal{O}_{n}|}{\Theta}\right),
    \label{eq:ienkf_1}
\end{equation} 
where $\mathcal{O}(t_n,i)$ is the predicted value for the observed quantities at time $t_n$ for particle $i$, and $\Theta$ is a ``temperature". In our case, $\mathcal{O}(t_n,i)$ is the prediction for the cumulative number of daily reported deaths $D_r(t_n,i)$. The filtering process is accomplished by keeping the particles with the largest weights with probability $\mathcal{P} = W(t_n, i)/\sum_j W(t_n, j)$. The states of the filtered particles will set the posterior distribution at time $t_n$, $X_F(t_n, i) = X_P(t_n, i_{best})$, where $i_{best}$ is the index of the filtered particles~\cite{king2008}. The parameter vector is updated at time $t_n$ using $\theta(t_{n},i) = \mathcal{N}(\theta(t_{n-1},i_{best}), b^{(m-1)}\sigma_{\theta})$. This filtering process continues until all the observations $\mathcal{O}_1,\dots,\mathcal{O}_N$ are compared. The iterative process continues by setting the initial state vector $X^{(m)}$ and parameter vector $\theta^{(m)}$ for the next iteration. The next parameter vector is given by~\cite{ionides2006}:
\begin{equation}
    \theta^{(m)} = \theta^{(m-1)} + V(t_1) \sum_{n=1}^{N} V^{-1}(t_n)\left(\bar{\theta}(t_{n}) - \bar{\theta}(t_{n-1})\right),
    \label{eq:ienkf_2}
\end{equation}
where $\bar{\theta}(t_{n})$ is the sample mean of $\theta(t_n, i_{best})$ and $V(t_n)$ is the variance~\cite{ionides2006, king2008}. The next state vector is given by the sample mean, 
\begin{equation}
  X^{(m)} = \frac{1}{P}\sum_{j_{best}=1}^P X(t_0,j_{best}).
  \label{eq:ienkf_3}
\end{equation}
After each iteration $m$, the initial state vector $X^{(m)}$ and parameter vector $\theta^{(m)}$ are used to compute the evolution of the model for the whole observation window $1,\dots,t_T$. The performance of the inferred model is computed by evaluating the error
\begin{equation}
  \varepsilon^{(m)}=\frac{1}{T} \sum_{n=1}^{T} \left| \mathcal{O}^{(m)}_n - \mathcal{O}_{n}\right|^2.
  \label{eq:ienkf_4}
\end{equation}
The iteration continues until $\left|\varepsilon^{(m)}-\varepsilon^{(m-1)}\right| <\epsilon_{max}$, where the threshold used here is $\epsilon_{max}=0.01$. In Table~S1 of the Supplementary Information, we report the values of the estimated parameters for consecutive time windows.

\section{Acknowledgments}

We gratefully acknowledge CNPq, CAPES, FUNCAP, the National Institute of Science and Technology for Complex Systems in Brazil and the Edson Queiroz Foundation for financial support.

\section{Contributions}

C.P., H.A.C., E.A.O., C.C., A.S.L., J.S.A. and V.F. designed research; C.P., H.A.C., E.A.O., C.C., A.S.L., J.S.A. and V.F. performed research; C.P., H.A.C., E.A.O., C.C., A.S.L., J.S.A. and V.F. analyzed data; and C.P., H.A.C., E.A.O., C.C., A.S.L., J.S.A. and V.F. wrote the paper. All authors reviewed the manuscript.

\section{Competing interests}

The authors declare no competing financial interests.

\section{Data Availability}


This study was approved by the Institutional Review Board (IRB) at Universidade de Fortaleza (UNIFOR). Two datasets were used with the approval and consent obtained by the Fortaleza City Hall, Ceará, Brazil. The first is a list of COVID-19 confirmed cases and deaths of patients in Fortaleza and the second consists of bus validations records from smart cards of passengers, both collected during the period from March to December, 2020. In the context of the ongoing health crisis, we make available these anonymized datasets under request to E. Oliveira at erneson@unifor.br.


\clearpage

\clearpage

\clearpage

\clearpage

\clearpage

\clearpage

\end{document}


\title{Supplementary Information: Tracing contacts to evaluate the transmission of COVID-19 from highly exposed individuals in public transportation}

\author{Caio Ponte$^{1}$,
Humberto A. Carmona$^{2}$,
Erneson A. Oliveira$^{1,3,4}\footnote{Correspondence to: erneson@unifor.br}$,
Carlos Caminha$^{1}$,
Antonio S. Lima Neto$^{5,6}$,
José S. Andrade Jr.$^{2}$,
Vasco Furtado$^{1,7}$}

\affiliation{$^1$ Programa de Pós Graduação em Informática Aplicada, Universidade de Fortaleza, 60811-905 Fortaleza, Ceará, Brasil.\\
$^2$ Departamento de Física, Universidade Federal do Ceará, 60455-760 Fortaleza, Ceará, Brasil.\\
$^3$ Laboratório de Ciência de Dados e Inteligência Artificial, Universidade de Fortaleza, 60811-905 Fortaleza, Ceará, Brasil.\\
$^4$ Mestrado Profissional em Ciências da Cidade, Universidade de Fortaleza, 60811-905, Fortaleza, Ceará, Brazil.\\
$^5$ Célula de Vigilância Epidemiológica, Secretaria Municipal da Saúde, 60810-670 Fortaleza, Ceará, Brasil.\\
$^6$ Departamento de Saúde Pública, Universidade de Fortaleza, 60451-970 Fortaleza, Ceará, Brasil.\\
$^7$ Empresa de Tecnologia da Informação do Ceará, Governo do Estado do Ceará, 60130-240 Fortaleza, Ceará, Brasil.}

\date{\today}

\maketitle

\section{Epidemiological curves for the compartmental model}

\begin{figure}[!htb]
\centering
\includegraphics[width=1.0\textwidth]{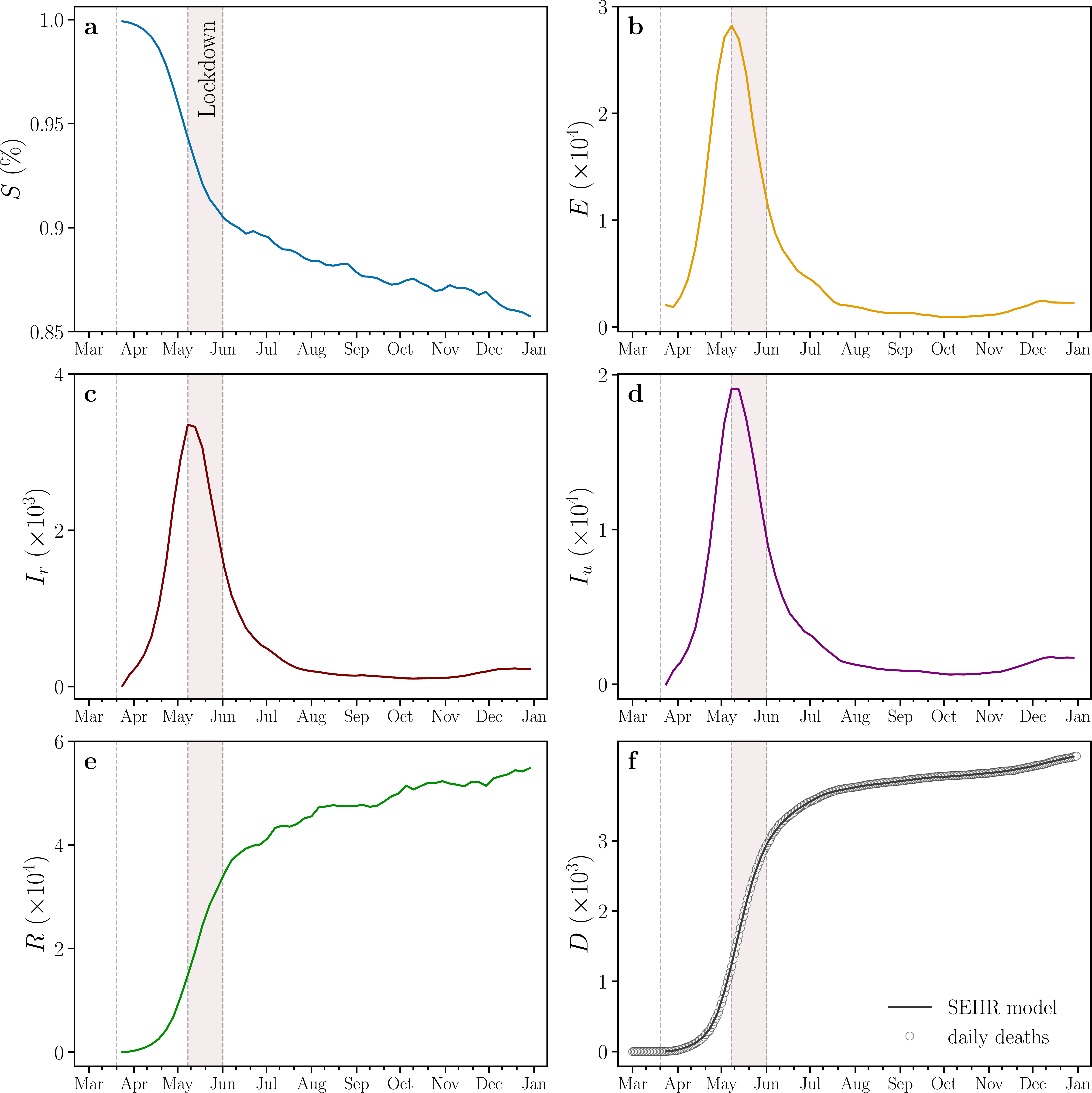}
\caption{{\bf Epidemiological curves obtained from the inference with the SEIIR  compartmental model.} Curves of the SEIIR model compartments. We show the curves with the values of the (a) susceptible, (b) exposed, (c) reported infectious, (d) unreported infectious, (e) recovered and (f) deceased population, respectively. The vertical dotted lines represent the beginning of social isolation (State Decree 33,519), lockdown (State Decree 33,574), and economic reopening (State Decree 33,608) regimes imposed on March 20, May 8, and June 1, 2020, respectively. We also highlight, in light red, the lockdown period in the city of Fortaleza.}
\label{fig:SEIIR_curves}
\end{figure}

We show all epidemiological curves for the proposed compartmental model in Fig.~\ref{fig:SEIIR_curves}a-\ref{fig:SEIIR_curves}f. The curve of the susceptible population is shown as a percentage. We estimate that about $20\%$ of the inhabitants of Fortaleza had contact with SARS-CoV-2 until December 2020. For the deceased population, we show the comparison between the observed cumulative number of deaths by SARS-CoV-2 in Fortaleza and our inferred model prediction.

\begin{figure}[!htb]
\centering
\includegraphics[width=1.0\textwidth]{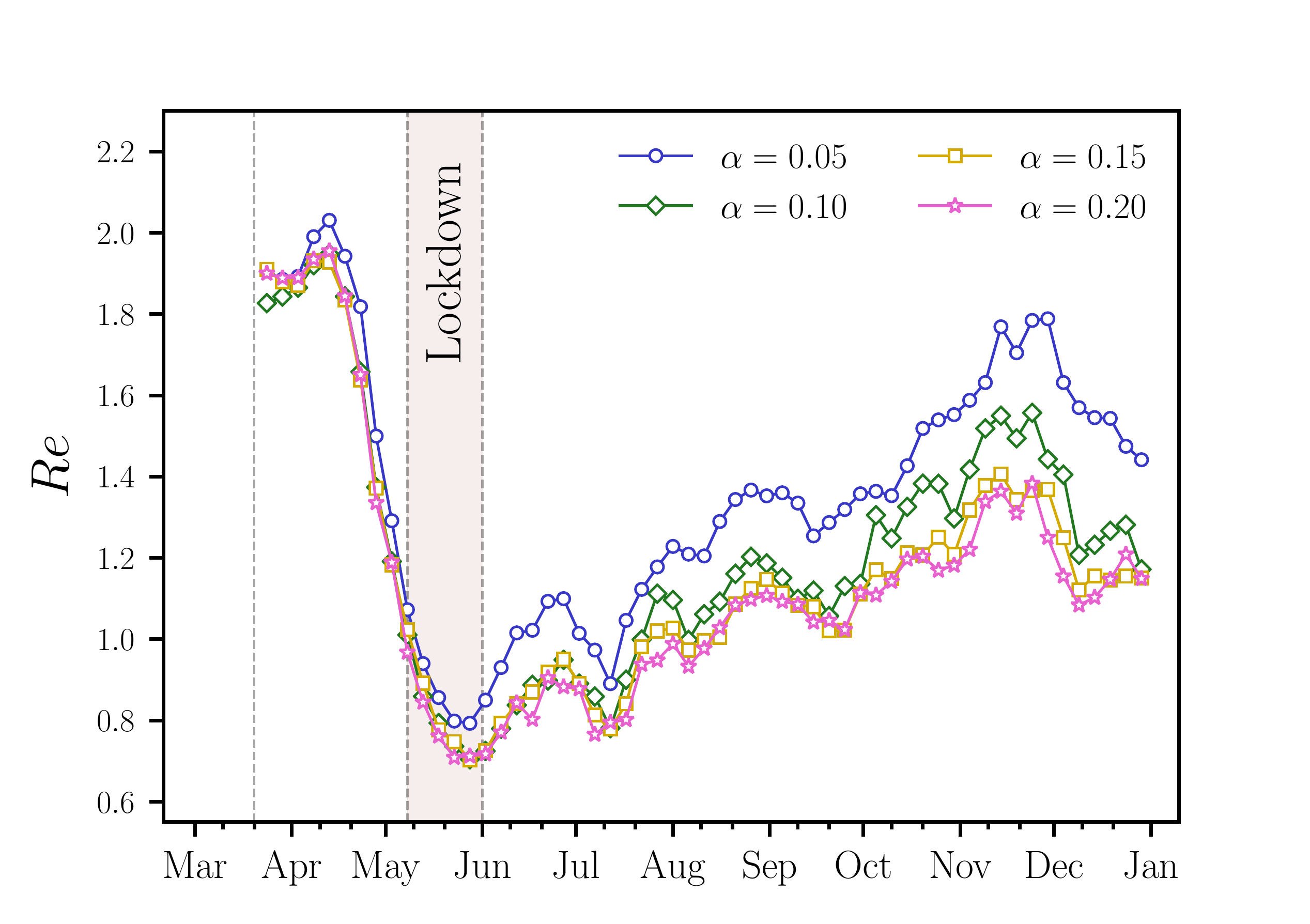}
\caption{{\bf Time Evolution of the effective reproduction number for the entire city}. We show the $Re^{city}$ curve computed from the SEIIR model for $\alpha = 0.055$, $0.10$, $0.15$ and $0.20$. Each point represents a bin of 22 days width and a 5 days step size. The vertical dotted lines represent the beginning of social isolation (State Decree 33,519), lockdown (State Decree 33,574), and economic reopening (State Decree 33,608) regimes imposed on March 20, May 8, and June 1, 2020, respectively. We also highlight, in light red, the lockdown period in the city of Fortaleza.}
\label{fig:SEIIR_Re}
\end{figure}

The time series of the effective reproduction number $Re^{city}$ for several values of $\alpha$ is shown in Fig.~\ref{fig:SEIIR_Re}. We find that $Re^{city}$ remains practically the same for $\alpha=0.15$ and $0.20$, which are compatible with the values of fractions of reported infectious individuals previously reported in the literature \cite{li2020b}. On March 20, 2020, the Government of Cear\'a imposed the first social distancing decree to its population, which led to a substantial decrease in human mobility everywhere in the state. However, as depicted, a value of $Re^{city}$ around $2$ in the beginning of April still indicated a rapid and exponential-like increase of the epidemic in the city of Fortaleza. A steep decrease in $Re^{city}$ was then observed and that effective reproduction number dropped below the critical value $1$ around May 5, 2020, when the Government decreed lockdown. The regime of lockdown imposed to the population of Fortaleza, in which only the commuting of essential workers were allowed, was responsible to keep the virus transmission under control ($Re^{city}<1$) until at least October 2020.

\section{Time Evolution of the upscaling factor}

\begin{figure}[!htb]
\centering
\includegraphics[width=1.0\textwidth]{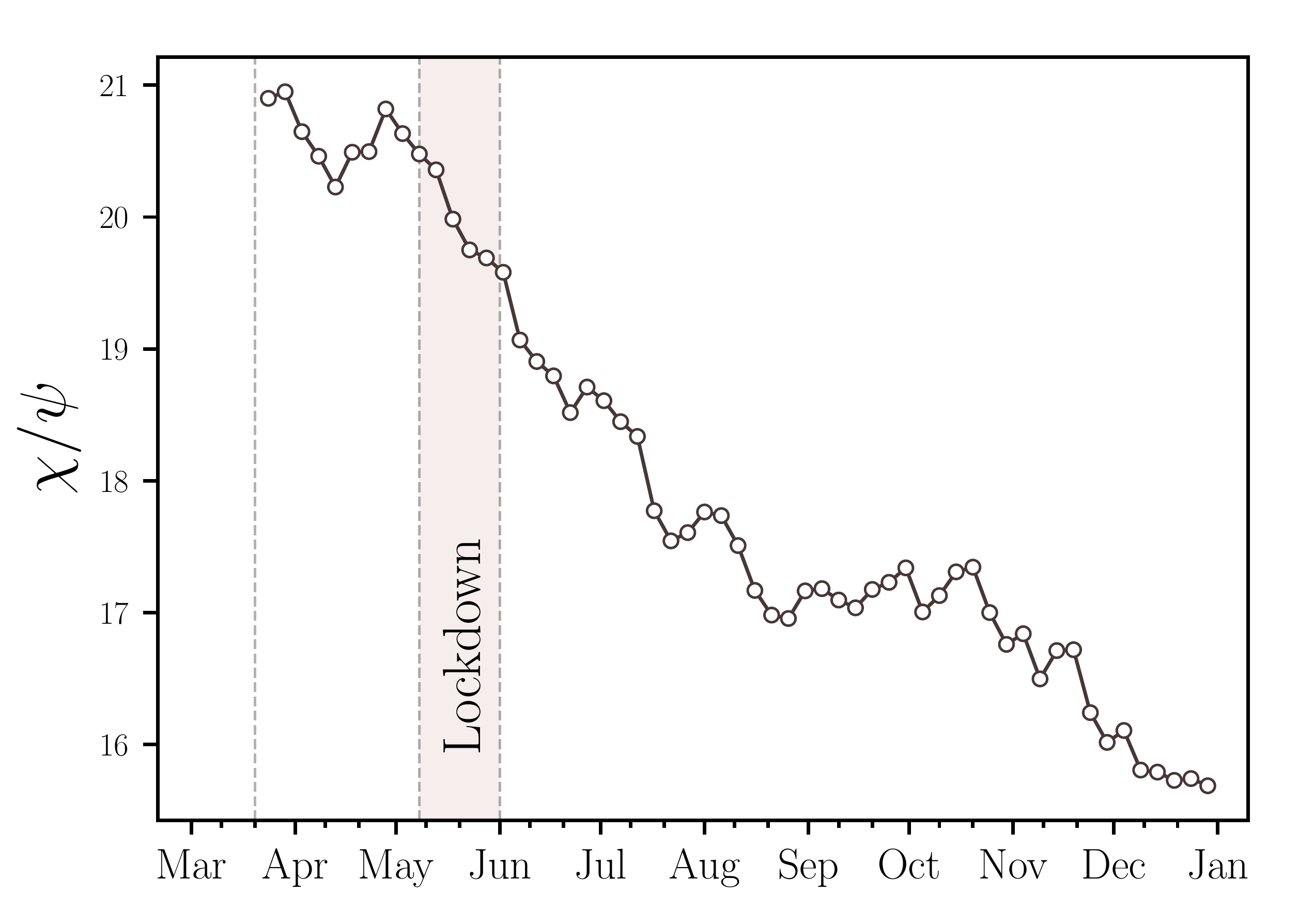}
\caption{{\bf Time Evolution of the upscaling factor}. The value of the upscaling factor $\chi/\psi$ computed from the SEIIR model decreases for consecutive time windows. Each point represents a bin of 22 days width and a 5 days step size. The vertical dotted lines represent the beginning of social isolation (State Decree 33,519), lockdown (State Decree 33,574), and economic reopening (State Decree 33,608) regimes imposed on March 20, May 8, and June 1, 2020, respectively. We also highlight, in light red, the lockdown period in the city of Fortaleza.}
\label{fig:factor}
\end{figure}

The time evolution of the upscaling factor $\chi/\psi$ is shown in Fig.~\ref{fig:factor}.

\section{Iterated Ensemble Kalman Filter Framework}

\begin{figure}[!htb]
\centering
\includegraphics[width=1.0\textwidth]{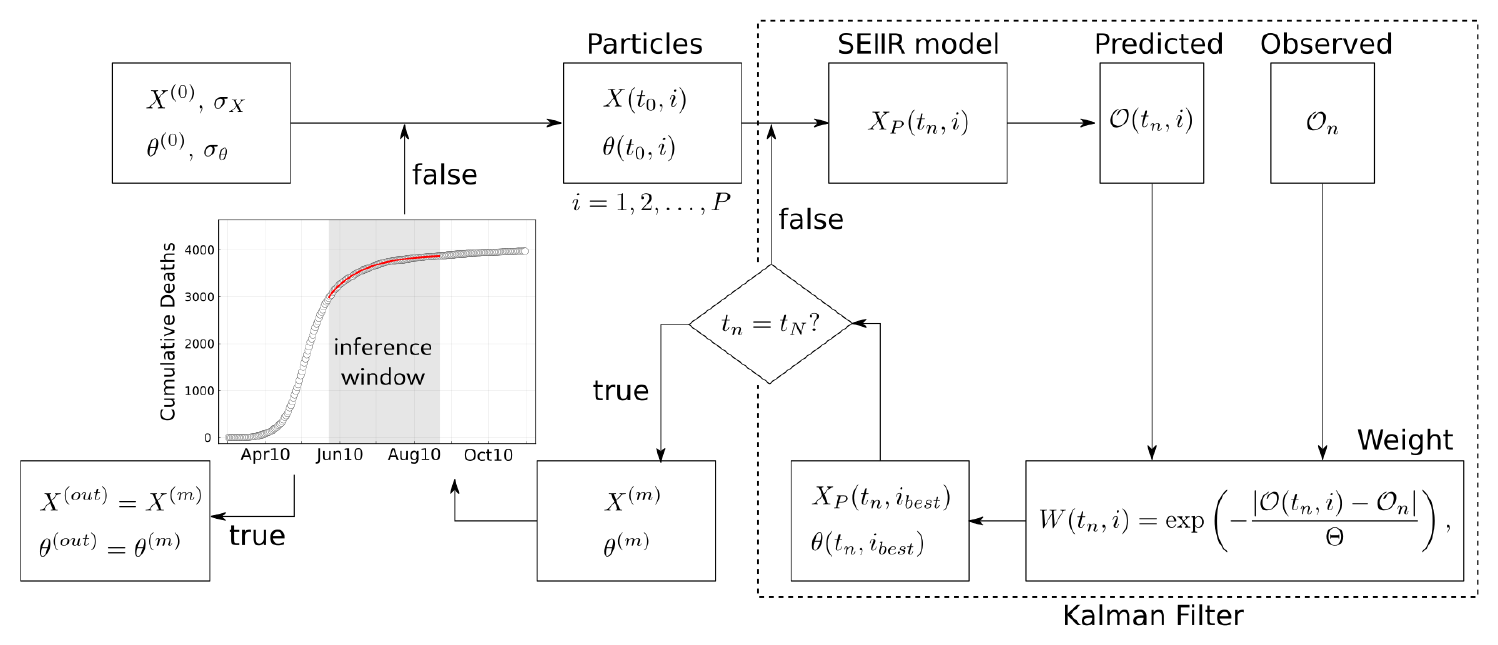}
\caption{{\bf Iterated Ensemble Kalman Filter (IEnKF) Framework.} Step-by-step flowchart of the IEnKF used for the inference with the compartmental model.}
\label{fig:framework}
\end{figure}

As shown in Fig.~\ref{fig:framework}, the flowchart of the Iterated Ensenble Kalman Filter (IEnKF)~\cite{li2020b, ionides2006, king2008, sakov2012} employed in our work includes the solution of the SEIIR model as a core modulus fed with initial guesses for state and parameter vectors. This modulus is built within the inference algorithm, from which estimated parameters and states during a given time window are obtained.

\section{Parameters of the compartmental model}

Table~\ref{tab:SEIIR_parameters} shows the epidemiological parameters of SEIIR model adopted for each window of inference.


\clearpage

\newpage

\newpage

\newpage

\newpage

\begin{longtable}{c|ccccccc}
\centering
Date & $\beta$ & $\mu$ & $\alpha$ & $\sigma$ & $\gamma$ & $\phi$ \\ \hline
2020-03-24 &  0.941742 &  0.486745 & 0.149977 & 0.223059 & 0.281187 & 0.069320 \\
2020-03-29 &  0.923219 &  0.486771 &  0.152547 &  0.221021 &  0.281529 &  0.069493 \\
2020-04-03 &  0.937314 &  0.476879 &  0.152980 &  0.221588 &  0.283996 &  0.067959 \\
2020-04-08 &  0.971974 &  0.471830 &  0.151557 &  0.218856 &  0.284957 &  0.066534 \\
2020-04-13 &  0.975671 &  0.464336 &  0.151683 &  0.222570 &  0.281936 &  0.066516 \\
2020-04-18 &  0.913568 &  0.471593 &  0.153497 &  0.220646 &  0.285345 &  0.067107 \\
2020-04-23 &  0.818039 &  0.472312 &  0.152583 &  0.225127 &  0.285158 &  0.071127 \\
2020-04-28 &  0.685570 &  0.480382 &  0.156198 &  0.231284 &  0.296786 &  0.075009 \\
2020-05-03 &  0.601887 &  0.472099 &  0.159905 &  0.232949 &  0.306464 &  0.076051 \\
2020-05-08 &  0.531756 &  0.467665 &  0.159058 &  0.232452 &  0.314061 &  0.078509 \\
2020-05-13 &  0.476636 &  0.464158 &  0.158516 &  0.234426 &  0.318556 &  0.077596 \\
2020-05-18 &  0.429962 &  0.452728 &  0.157680 &  0.235465 &  0.321578 &  0.078138 \\
2020-05-23 &  0.418717 &  0.445415 &  0.157436 &  0.232583 &  0.315319 &  0.076398 \\
2020-05-28 &  0.393294 &  0.444844 &  0.155333 &  0.230803 &  0.316501 &  0.076914 \\
2020-06-02 &  0.410878 &  0.443603 &  0.151952 &  0.228969 &  0.315920 &  0.076274 \\
2020-06-07 &  0.460817 &  0.428248 &  0.150497 &  0.232291 &  0.315902 &  0.075106 \\
2020-06-12 &  0.496968 &  0.424960 &  0.147721 &  0.227823 &  0.313617 &  0.074771 \\
2020-06-17 &  0.520876 &  0.423615 &  0.144601 &  0.228621 &  0.314275 &  0.073622 \\
2020-06-22 &  0.553313 &  0.415472 &  0.143676 &  0.222961 &  0.313110 &  0.073552 \\
2020-06-27 &  0.563638 &  0.422421 &  0.142414 &  0.220978 &  0.315451 &  0.071979 \\
2020-07-02 &  0.537569 &  0.420991 &  0.139717 &  0.220920 &  0.316575 &  0.071163 \\
2020-07-07 &  0.492326 &  0.414203 &  0.142366 &  0.221908 &  0.313338 &  0.069663 \\
2020-07-12 &  0.472757 &  0.410928 &  0.142010 &  0.226979 &  0.313961 &  0.067919 \\
2020-07-17 &  0.529636 &  0.394527 &  0.140153 &  0.224913 &  0.312099 &  0.067265 \\
2020-07-22 &  0.623454 &  0.389191 &  0.137581 &  0.221728 &  0.310518 &  0.065810 \\
2020-07-27 &  0.651591 &  0.391906 &  0.136511 &  0.224862 &  0.309801 &  0.065467 \\
2020-08-01 &  0.646985 &  0.394598 &  0.139634 &  0.229596 &  0.310438 &  0.066236 \\
2020-08-06 &  0.619746 &  0.394033 &  0.139218 &  0.229674 &  0.312464 &  0.065691 \\
2020-08-11 &  0.636491 &  0.386854 &  0.139302 &  0.230407 &  0.309448 &  0.065631 \\
2020-08-16 &  0.649350 &  0.374605 &  0.141486 &  0.231604 &  0.308376 &  0.066573 \\
2020-08-21 &  0.700571 &  0.369588 &  0.140327 &  0.231880 &  0.302333 &  0.067226 \\
2020-08-26 &  0.728535 &  0.368281 &  0.140994 &  0.230835 &  0.301127 &  0.066636 \\
2020-08-31 &  0.737557 &  0.374409 &  0.141605 &  0.228592 &  0.296083 &  0.067373 \\
2020-09-05 &  0.706375 &  0.375461 &  0.140895 &  0.226542 &  0.295648 &  0.067289 \\
2020-09-10 &  0.691229 &  0.374577 &  0.138387 &  0.226096 &  0.295912 &  0.067328 \\
2020-09-15 &  0.690880 &  0.372604 &  0.138551 &  0.227870 &  0.291531 &  0.067811 \\
2020-09-20 &  0.648453 &  0.375282 &  0.140883 &  0.232778 &  0.289339 &  0.066958 \\
2020-09-25 &  0.650962 &  0.375601 &  0.142775 &  0.231963 &  0.292576 &  0.067333 \\
2020-09-30 &  0.705538 &  0.378565 &  0.143436 &  0.231854 &  0.298863 &  0.066841 \\
2020-10-05 &  0.758124 &  0.370228 &  0.140453 &  0.233296 &  0.298339 &  0.067408 \\
2020-10-10 &  0.742961 &  0.374269 &  0.140271 &  0.229163 &  0.296711 &  0.066331 \\
2020-10-15 &  0.766765 &  0.380999 &  0.138758 &  0.229484 &  0.292138 &  0.066040 \\
2020-10-20 &  0.763361 &  0.381141 &  0.140105 &  0.228106 &  0.297784 &  0.067130 \\
2020-10-25 &  0.802157 &  0.370811 &  0.139463 &  0.228721 &  0.302356 &  0.067226 \\
2020-10-30 &  0.782503 &  0.361956 &  0.141222 &  0.225585 &  0.304570 &  0.066343 \\
2020-11-04 &  0.845170 &  0.365497 &  0.139852 &  0.229970 &  0.301449 &  0.065995 \\
2020-11-09 &  0.911679 &  0.356743 &  0.137196 &  0.232050 &  0.308682 &  0.064187 \\
\caption{{\bf Parameters of the compartmental model.} The epidemiological parameters of SEIIR model adopted for each window of inference.}
\label{tab:SEIIR_parameters}
\end{longtable}
\newpage